\documentclass[aps,showpacs,prl,twocolumn,footinbib,superscriptaddress]{revtex4-1}
\usepackage{epsfig}
\usepackage{amsfonts}
\usepackage{amsmath}
\usepackage{bm}
\usepackage{xcolor}
\usepackage{newtxtext,newtxmath}
\usepackage{booktabs}
\usepackage{yfonts}

\usepackage[utf8]{inputenc}
\usepackage[colorlinks,bookmarks=false,citecolor=blue,linkcolor=blue,urlcolor=blue]{hyperref}
\usepackage{soul}
\usepackage{stackengine} 

\usepackage{changepage}

\newcommand{\bra}[1]{\left< #1 \right|}
\newcommand{\ket}[1]{\left| #1 \right>}

\newcommand{\ketbra}[2]{\left| #1 \right> \left< #2 \right|}

\newcommand{\average}[3]{\left< #1 | #2 | #3 \right>}

\newcommand{\be}{\begin{equation}}
\newcommand{\ee}{\end{equation}}

\begin{document}

\title{Exploring the many-body dynamics near a conical intersection with trapped Rydberg ions}
\author{Filippo M. Gambetta}
\affiliation{School of Physics and Astronomy, University of Nottingham, Nottingham, NG7 2RD, United Kingdom}
\affiliation{Centre for the Mathematics and Theoretical Physics of Quantum Non-equilibrium Systems, University of Nottingham, Nottingham NG7 2RD,  United Kingdom}
\author{Chi Zhang}
\affiliation{Department of Physics, Stockholm University, 10691 Stockholm, Sweden}
\author{Markus Hennrich}
\affiliation{Department of Physics, Stockholm University, 10691 Stockholm, Sweden}
\author{Igor Lesanovsky}
\affiliation{School of Physics and Astronomy, University of Nottingham, Nottingham, NG7 2RD, United Kingdom}
\affiliation{Centre for the Mathematics and Theoretical Physics of Quantum Non-equilibrium Systems, University of Nottingham, Nottingham NG7 2RD,  United Kingdom}
\affiliation{Institut für Theoretische Physik, University of Tübingen, 72076 Tübingen, Germany}
\author{Weibin Li}
\affiliation{School of Physics and Astronomy, University of Nottingham, Nottingham, NG7 2RD, United Kingdom}
\affiliation{Centre for the Mathematics and Theoretical Physics of Quantum Non-equilibrium Systems, University of Nottingham, Nottingham NG7 2RD,  United Kingdom}

\date{\today}

\begin{abstract}
Conical intersections between electronic potential energy surfaces are paradigmatic for the study of nonadiabatic processes in the excited states of large molecules. However, since the corresponding dynamics occurs on a femtosecond timescale, their investigation remains challenging and requires ultrafast spectroscopy techniques. We demonstrate that trapped Rydberg ions are a platform to engineer conical intersections and to simulate their ensuing dynamics on larger length and time scales of the order of nanometers and microseconds, respectively; all this in a highly controllable system. Here, the shape of the potential energy surfaces and the position of the conical intersection can be tuned thanks to the interplay between the high polarizability and the strong dipolar exchange interactions of Rydberg ions. We study how the presence of a conical intersection affects both the nuclear and electronic dynamics demonstrating, in particular, how it results in the inhibition of the nuclear motion. These effects can be monitored in real-time via a direct spectroscopic measurement of the electronic populations in a state-of-the-art experimental setup. 
\end{abstract}

\maketitle

\textit{Introduction.---} The Born-Oppenheimer (BO) approximation is a cornerstone of modern solid state and molecular theories~\cite{Born:1927, Domcke:2004}. It provides a simple description of many electronic systems in which nuclei move on a single potential energy surface (PES) generated by the electronic dynamics. Nevertheless, nonadiabatic phenomena, which fall outside the validity regime of the BO approximation, are ubiquitous in many fundamental chemical processes~\cite{Tully:2012}. Among them, conical intersections (CIs), which occur when two (or more) PESs are degenerate within a given sub-manifold of the nuclear coordinates, are at basis of many fundamental photo-chemical processes in large molecules~\cite{Yarkony:1996,Domcke:2004,Baer:2006}. They provide a fast and radiationless de-excitation mechanism between intra-molecular electronic states~\cite{Ismail:2002,Perun:2005}, which contributes to the stability of DNA~\cite{Barbatti:2010}, to the mechanism of vision ~\cite{Schoenlein:1991,Polli:2010, Rinaldi:2014}, and to photosynthesis~\cite{Hammarstrom:2008}. Importantly, CIs are tightly linked to the emergence of geometric phase (GP) effects: during the motion on a path that encircles a CI, both the electronic and nuclear wavefunctions acquire an extra phase factor of $ \pi $, which results in observable interference phenomena~\cite{Bohm:2003, Longuet-Higgins:1958, Berry:1984, Mead:1992,Althorpe:2008, Ryabinkin:2013, Ryabinkin:2017}. 
Despite the understanding of CIs developed over the past decades, passages through a CI in real time have been observed only recently~\cite{Yarkony:1996,Domcke:2004, Baer:2006, Meyer:2009,Polli:2010}. Since processes associated with CIs occur typically on a femtosecond timescale, ultrafast and broadband spectroscopy techniques are indeed needed to record the features  of dynamics they induce~\cite{Nunn:2010, Kowalewski:2015, Adachi:2019, Young:2018}. Moreover, the observation of GP effects requires a high degree of control on the preparation of the nuclear wavefunction, which is hardly achievable in real molecules. In recent years, ultracold gases have been proposed as a tool to overcome all of these issues and to study CIs in a well-controlled environment~\cite{Sindelka:2011, Moiseyev:2008, Wallis:2009}, with Rydberg atoms~\cite{Gallagher:2005,Low:2012} suggested as a quantum simulator for the dynamics in the neighborhood of a CI~\cite{Wuster:2011, Wuster:2018}. 

\begin{figure}[t]
	\centering
	\includegraphics[width=\columnwidth]{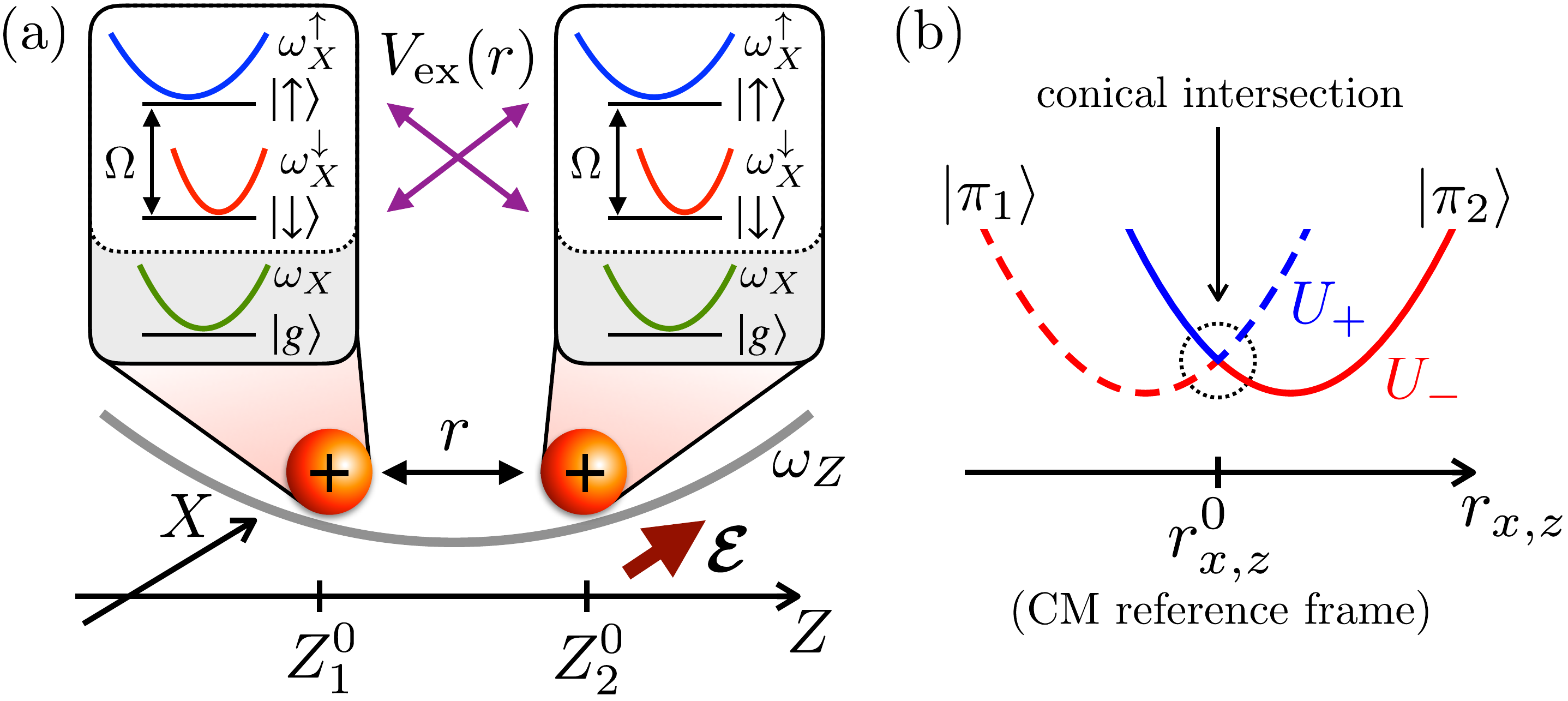}
	\caption{{\bf Conical intersection in a system of two Rydberg ions.} (a) Two ions are confined by a harmonic trapping potential in the $ X- Z $ plane. From the ground state $\ket{g} $, each ion is excited to the two Rydberg states $ \ket{\downarrow}=\ket{nS}  $ and $ \ket{\uparrow}=\ket{nP} $. Due to different polarizabilities, ions in Rydberg states experience a state-dependent trapping potential along $ X $, with frequencies $ \omega^{\downarrow,\uparrow}_X $, and different equilibrium positions. When excited to the pair states $ \ket{\pi_1}=\ket{\uparrow \downarrow} $ and $ \ket{\pi_2}=\ket{\downarrow \uparrow} $, ions interact through the exchange interaction $ V_\mathrm{ex}$, whose strength depends on their separation $ r $. A static offset electric field $ \bm{\mathcal{E}} $ allows to control the transverse equilibrium position of the ions. (b) The interplay between the exchange interaction and the state-dependent confinement shifts the energies of the states $ \ket{\pi_1} $ (dashed curve) and $ \ket{\pi_2} $ (solid curve). This results in a conical intersection in the potential energy surfaces $ U_- $ (red) and $ U_+ $ (blue) which, in the reference frame of the center of mass of the system, occurs at the ions' relative equilibrium separation $ r_{x,z}^0 $. See text for details.
	\label{fig:setup}}
\end{figure}

In this work we show that trapped Rydberg ions offer ideal properties for controlling and investigating the quantum many-body dynamics near a CI. They combine the high degree of control typical of trapped ion setups with tunable dipole-dipole interactions, enabled by the
possibility to individually excite each ion to a high-lying Rydberg level~\cite{Muller:2008,Schmidt-Kaler:2011,Li:2012,Li:2014,Higgins:2017,Higgins:2017PRL,Mokhberi:2019,Vogel:2019,Gambetta:2020b,Zhang:2020}.  
Moreover, due to the interplay between the state-dependent polarizability of Rydberg states and the radiofrequency electric field of the Paul trap, Rydberg ions feature a controllable state-dependent trapping potential~\cite{Higgins:2019,Higgins:2019:thesis}. 
We show that this mechanism, in the presence of two different Rydberg states and combined with strong dipolar exchange interactions between two Rydberg ions, can be exploited to realize a minimal instance of a CI, consisting of two electronic states and two non-trivial nuclear coordinates~\cite{Domcke:2004}, shown in Fig.~\ref{fig:setup}. The PESs and the position of the CI are controlled by the exchange interactions and an external electric field, allowing one to realize various scenarios which occur, e.g., in photo-chemical processes, but on nanometer length and microsecond time scales. The high degree of control over the vibrational motion of the ions makes it possible to minimize decoherence effects stemming from the coupling to additional rovibrational nuclear degrees of freedom. The latter, which cannot be avoided in real molecules, affects the dynamics across a CI on a sub-picoseconds timescale and hinders the experimental observation of coherent quantum effects~\cite{Kuhl:2001,Chen:2016,Ulbricht:2016}. To demonstrate the capabilities of our approach, we investigate effects induced by the GP in a feasible experimental setup: We show that the destructive interference between the paths encircling the different sides of a CI results in the localization of the nuclear wavepacket. This GP effect can be directly observed via a measurement of the electronic state population and does not require the spatially resolved detection of each ion. 

\textit{Equations of motion.---}
We consider a system of two Rydberg ions of mass $ m $ in a linear Paul trap. The latter gives rise to an effective harmonic trapping potential with frequencies $ \omega_{X,Y,Z} $~\cite{Major:2005,Higgins:2019,Higgins:2019:thesis}. For the sake of simplicity, we assume $ \omega_Y\gg\omega_X>\omega_Z $, so that the motion of the ions is confined to the $ X-Z $ plane, with the $ Z- $axis being the longitudinal one; see Fig.~\ref{fig:setup}. The potential energy of the two trapped ions is $ V_\mathrm{trap}=m [\omega_X^2 (X_1^2+X_2^2)+ \omega_Z^2 (Z_1^2+Z_2^2)]/2+ ke^2/r $, where the last term corresponds to the repulsive Coulomb interaction [with $ k=1/(4\pi\epsilon_0) $ being the Coulomb constant]. Here, $ \bm{r}=\bm{R}_1-\bm{R}_2 $, with $ \bm{R}_i=(X_i,Z_i) $, $ i\in\{1,2\} $, the nuclear coordinates in the laboratory frame. From their ground state $ \ket{g} $, the ions can be excited to the two Rydberg levels $ \ket{nS}=\ket{\downarrow} $ and $ \ket{nP}= \ket{\uparrow} $, with $ n $ being the principal quantum number, both of which can be coupled by a microwave (MW) field with Rabi frequency $ \Omega $. Rydberg-excited ions interact via the exchange interaction potential $ V_\mathrm{ex}(r)\cdot(\sigma_+^1\otimes\sigma_-^2+\sigma_-^1\otimes\sigma_+^2) $, with $ \sigma_\pm^i $ acting on the $ i-$th ion as $ \sigma^i_+\ket{\downarrow^i}=\ket{\uparrow^i}$ and $ \sigma^i_-\ket{\uparrow^i}=\ket{\downarrow^i} $. Furthermore, due to the large polarizability of Rydberg levels (denoted by $ \rho_\sigma $, with $ \sigma\in\{\downarrow,\uparrow\} $), ions in the Rydberg states experience an additional transverse trapping potential $ \delta V_{\mathrm{trap},\sigma} = -\rho_\sigma A^2 (X_1^2+X_2^2)$, where $ A $ is the gradient of the radio-frequency field of the Paul trap~\cite{Higgins:2019, Higgins:2019:thesis}.  Finally, the transverse equilibrium positions of the ions is controlled via a static offset electric field $ \bm{\mathcal{E}} $ along the $ X- $axis, which results in the potential energy contribution$ V_\mathrm{mm}=e\mathcal{E}(X_1+X_2) $. In what follows we will focus on the dynamics in the spin subspace $ \mathcal{H}_\mathrm{sp} $, which is spanned by the ion pair states $ \ket{\pi_1}=\ket{\uparrow\downarrow} $ and $ \ket{\pi_2}=\ket{\downarrow\uparrow} $. The latter define the so-called \emph{diabatic basis} of the system~\cite{Wuster:2011, Wuster:2018}. In this subspace, we define the Pauli operators $ S_0=\ketbra{\pi_1}{\pi_1}+\ketbra{\pi_2}{\pi_2} $, $ S_x=\ketbra{\pi_1}{\pi_2}+\ketbra{\pi_2}{\pi_1} $, and $ S_z=\ketbra{\pi_1}{\pi_1}-\ketbra{\pi_2}{\pi_2} $. 

The system is most conveniently analyzed in the center of mass (CM) reference frame, with $ \bm{R}=(X,Z)= (\bm{R}_1+\bm{R}_2)/2 $ and $ \bm{r}=(x,z) $ the CM and the relative coordinates, respectively.  In terms of the latter, the full system Hamiltonian is
\begin{equation}\label{eq:Hfull}
H = \left(-\frac{\nabla^2_{\bm{R}}}{2M}-\frac{\nabla^2_{\bm{r}}}{2\mu}\right) \otimes S_0 + H_\mathrm{spin},
\end{equation}
where the first term accounts for the total kinetic energy, with $ M=2m $ and $ \mu=m/2 $ the CM and reduced mass, respectively. The potential energy contributions are contained in $ H_\mathrm{spin}=H_\mathrm{spin}^0 +  H_\mathrm{spin}^1$, with
\begin{subequations}
\begin{align}\label{eq:H0andH1}
H_\mathrm{spin}^0&=\left(V_\mathrm{CM}+V_\mathrm{rel}\right)\otimes S_0,\\
H_\mathrm{spin}^1&=V_\mathrm{ex}(r)\otimes S_x + H_\mathrm{CM-rel}.\label{eq:Hspin1}
\end{align}
\end{subequations}
Here, by defining the polarizabilities $ \rho_\pm=\rho_\uparrow\pm\rho_\downarrow $, we have
\begin{subequations}\label{eq:2ions:HtermsCoMsub}
	\begin{align}
	V_\mathrm{CM}&=\frac{M}{2}\left[\bar{\omega}_X^2 \left(X-X^0\right)^2+\omega_Z^2 Z^2\right], \label{eq:2ions:HtermsCoMsub:a}\\
    V_\mathrm{rel}&=\frac{\mu}{2}\left(\bar{\omega}_X^2 x^2+\omega_Z^2 z^2\right)+\frac{ke^2}{r}.\label{eq:2ions:HtermsCoMsub:b}
	\end{align}
\end{subequations}
Equation~\eqref{eq:2ions:HtermsCoMsub:a} corresponds to a harmonic trapping potential with renormalized frequency $ \bar{\omega}_X^2=\omega_X^2 - A^2\rho_+/m $ and displaced along the $ x- $direction by $ \bm{R}^0=(X^0,0) $, where $ X^0=-e\mathcal{E}/(m\bar{\omega}_X^2)  $ is controlled by the external electric field $ \mathcal{E} $. On the other hand,  Eq.~\eqref{eq:2ions:HtermsCoMsub:b} contains the competition between the harmonic potential and the repulsive Coulomb interaction. Finally, the last term in Eq.~\eqref{eq:Hspin1} is given by $ H_\mathrm{CM-rel}=A^2 \rho_- Xx \otimes S_z $ and describes the coupling between the electronic and nuclear motion arising from the unequal polarizabilities of the Rydberg states.

For typical experimental values, the Coulomb repulsion represents the largest energy scale of the system associated with mechanical motion. Therefore, to investigate its dynamics, one can treat $ H_\mathrm{spin}^1 $ as a small perturbation to $ H_\mathrm{spin}^0 $~\cite{Note2}. We first perform a harmonic approximation around the unperturbed equilibrium positions of the ions, $ \bm{R}^0 $ and $ \bm{r}^0 $, by introducing the displacements $ \bm{Q}=\bm{R}-\bm{R}^0 $ and $ \bm{q}=\bm{r}-\bm{r}^0 $. Here, $ \bm{R}^0 $ and $  \bm{r}^0 $ are determined by Eq.~\eqref{eq:2ions:HtermsCoMsub}. In particular, $ \bm{r}^0=(0,z^0) $, where $ x^0=0 $ due to symmetry considerations and $ z^0 $ is given by the solution of $ \partial_z V_\mathrm{rel}|_{z=z_0}=0 $. We can then expand $ H_\mathrm{spin}^1 $ to the first order in the displacements $ \bm{Q} $ and $ \bm{q} $ and obtain the full spin Hamiltonian $ H_\mathrm{spin} $ in the harmonic approximation. An examination of the various contributions to Eq.~\eqref{eq:H0andH1} suggests that the most interesting part of the dynamics occurs in the relative coordinate sector, as we show in the Supplemental Material (SM)~\cite{Note1}. By neglecting the motion of the CM (i.e., by setting $ \bm{Q}=\bm{0} $), and focusing exclusively on the relative coordinate $ \bm{q} $, we finally obtain 
\begin{equation}\label{eq:HspinSGW}
H_\mathrm{spin} \approx S(\bm{q}) \otimes S_0 + G(\bm{q}) \otimes S_z + W(\bm{q}) \otimes S_x,
\end{equation}
where $ S(\bm{q})=  \mu \bm{q}^T \mathcal{K}^q \bm{q}/2 $, and
\begin{subequations}\label{eq:SGW1}
	\begin{align}
	G(\bm{q})&= A^2 \rho_- X^0 q_x,\\
	W(\bm{q})&= V^0_\mathrm{ex}+F_z^0 q_z.\label{eq:WQ0}
	\end{align}
\end{subequations}
Here, we have introduced the dynamical matrix $ \mathcal{K}^q_{\alpha \beta}=\mu^{-1}\partial_{r_\alpha,r_\beta} V_\mathrm{rel}|_{\bm{R}^0,\bm{r}^0} $, with $ \alpha,\beta\in\{x,z\} $, defined $ V^0_\mathrm{ex}=V_\mathrm{ex}(r^0) $, and denoted with $ F_z^0 $ the $ z- $component of $ \nabla_{\bm{r}}V_\mathrm{ex}(r)|_{\bm{r}=\bm{r}^0} $.

In the Hamiltonian of Eq.~\eqref{eq:HspinSGW} one can recognize the minimal model which can host a CI~\cite{Ryabinkin:2017, Domcke:2004}. This represents the main result of our work and shows that a CI can be actually realized in an experimentally feasible system of two trapped Rydberg ions. The high degree of control on both electronic and vibrational degrees of freedom makes this platform an ideal candidate for the study of the coherent dynamics occurring in neighborhood of a CI, as we will show in next sections.

\begin{figure}[t]
	\centering
	\includegraphics[width=\columnwidth]{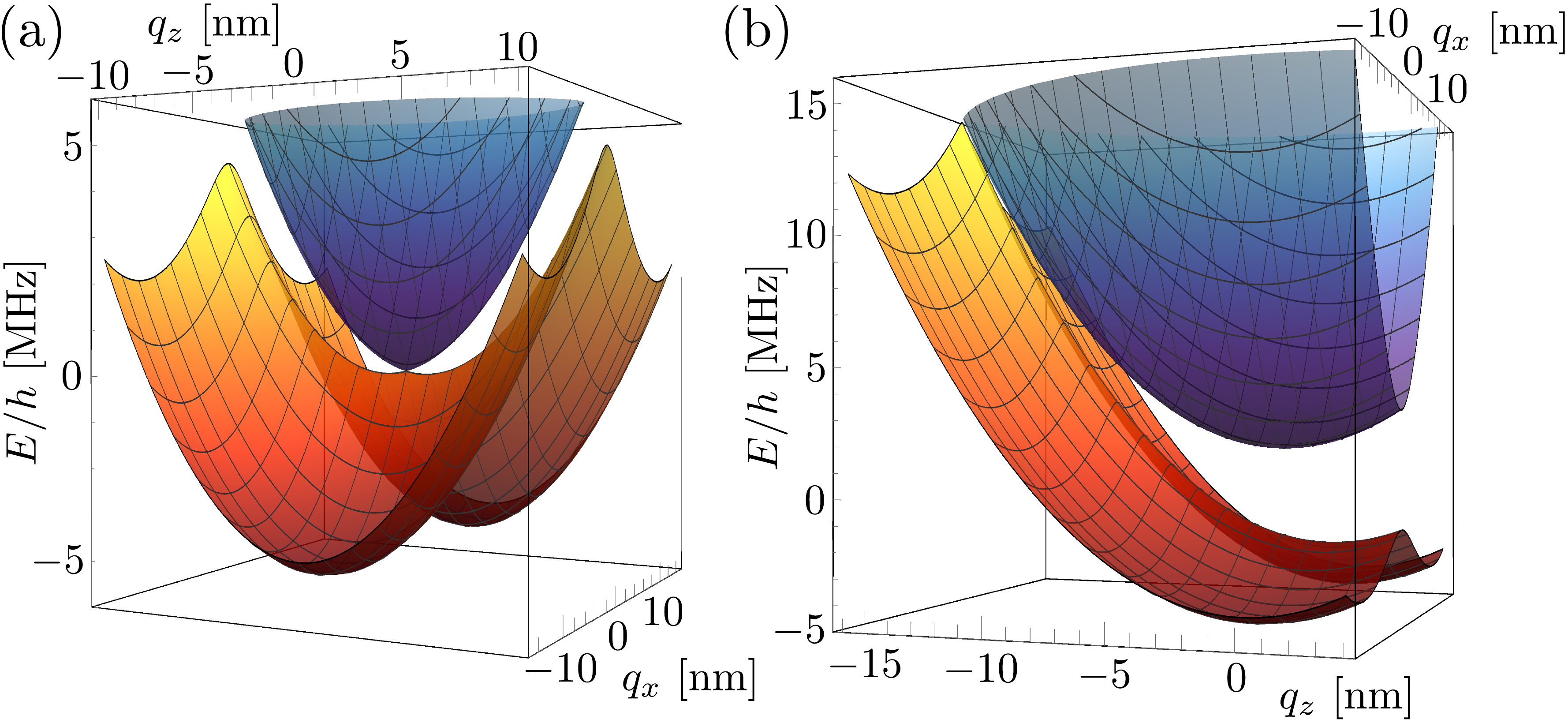}
	\caption{{\bf Potential energy surfaces in the branching plane.} Plot of the eigenvalues of Eq.~\eqref{eq:HspinSGW} (i.e., the PESs) $ U_\pm(\bm{q}) $ as a function of the displacements $ q_x $ and $ q_z $. (a) For $ V^0_\mathrm{ex}/h=0 $, a CI occurs at $ \bm{q}^* =(0,0) $ while for (b) $ V^0_\mathrm{ex}/h=2\pi \times 0.3 $ MHz the CI is shifted to $ \bm{q}^* =(0,-V^0_\mathrm{ex}/F_z^0 ) $ (see text for details). Here, we considered $ {}^{88}\mathrm{Sr}^+ $ Rydberg ions, with $ m=87.9\times 1.66 \times 10^{-27}\ \mathrm{kg}$, and set $ \omega_X = 2\pi\times1.6\ \mathrm{MHz}$, $ \omega_Z = 2\pi\ \mathrm{MHz} $, $ \mathcal{E}=2.529\ \mathrm{V/m} $ (resulting in $ X^0=-0.024\ \mu\mathrm{m} $), $ \rho_\downarrow=8.9 \times 10^{-31}\ \mathrm{C^2m^2/J} $, $ \rho_\uparrow=-3.8 \times 10^{-30}\ \mathrm{C^2m^2/J}$ (corresponding to Rydberg $ \ket{nS} $ and $ \ket{nP} $ states with $ n=50 $), and $ F_z^0/h =2\pi\times 20 \ \mathrm{MHz/\mu m}$. \label{fig:PES}}
\end{figure}

\textit{CI-induced dynamics in the nuclear motion.---} 
In the spirit of the Born-Huang approach~\cite{Born:1954}, the nuclear motion of the ions is determined by the PESs. The latter are given by the eigenvalues of the electronic Hamiltonian $ H_\mathrm{spin} $,
\begin{equation}\label{eq:Hspineigval}
U_\pm(\bm{q}) = S(\bm{q}) \pm \sqrt{[G(\bm{q})]^2+[W(\bm{q})]^2}.
\end{equation} 
In general, CIs arise at positions $ \bm{q}^* $ such that $ U_+(\bm{q}^*)=U_-(\bm{q}^*) $. From Eq.~\eqref{eq:Hspineigval} this is equivalent to $G(\bm{q}^*)=W(\bm{q}^*)=0 $, which implies that the system described by Eq.~\eqref{eq:HspinSGW} has a CI at $ q^*_x=0$ and $q^*_z=-V^0_\mathrm{ex}/F_z^0  $. Two examples of the PESs in the neighborhood of the latter are shown in Fig.~\ref{fig:PES} for different values of $ V^0_\mathrm{ex} $. 

The eigenstates associated with the PESs $ U_\pm(\bm{q}) $, which define the electronic \emph{adiabatic basis}, are~\cite{Wuster:2011, Wuster:2018}
\begin{subequations}\label{eq:adiabaticbasis}
	\begin{align}\label{key}
	\ket{\varphi_+(\bm{q})} &= \cos[\Lambda(\bm{q})] \ket{\pi_1} + \sin[\Lambda(\bm{q})] \ket{\pi_2}, \\
	\ket{\varphi_-(\bm{q})} &= - \sin[\Lambda(\bm{q})] \ket{\pi_1} + \cos[\Lambda(\bm{q})] \ket{\pi_2}, 
	\end{align}
\end{subequations}
where $ \Lambda(\bm{q}) $ is fixed by $ \tan [2\Lambda(\bm{q})] = W(\bm{q})/G(\bm{q}) $. Note that if $ \bm{q} $ is varied along a close path encircling the CI, the mixing angle $ \Lambda(\bm{q}) $ changes only by $ \pi $ instead of $ 2\pi $: the states $ \ket{\varphi_\pm(\bm{q})} $ acquire an extra phase of $ \pi $, known as the GP~\cite{Ryabinkin:2017,Berry:1984,Longuet-Higgins:1958}.

For the sake of simplicity, we first consider a suitably tailored exchange interaction potential such that $ V^0_\mathrm{ex}=0 $~\cite{Gambetta:2020b, Note1}. In this case, the CI occurs at $ \bm{q}^*=\bm{0} $ and the PESs, shown in Fig.~\ref{fig:PES}(a), are symmetric under the reflections $ \bm{q} \rightarrow -\bm{q} $. As a consequence, the two paths connecting the two minima of $ U_-(\bm{q}) $ and encircling the CI from opposite sides are identical. This represents the ideal setting to investigate dynamical effects induced by the presence of a CI and, in particular, those related to the GP. 

To do so, we first initialize the system's nuclear wavefunction in the state $ \phi_{\mathrm{rel}}^\mathrm{ss}(\bm{q}) $, corresponding to a Gaussian centered around $ (q^0_{x,\mathrm{ss}},0) $~\cite{Note1}. Here, we set $ q^0_{x,\mathrm{ss}} $ to coincide with one of the two minima of $ U_-(\bm{q}) $ [see Fig.~\ref{fig:PES}(a)]. As explained in the SM~\cite{Note1}, this allows us to maximize the visibility of the GP effects thanks to symmetries of the system.
Then, to initiate the dynamics in the spin subspace $ \mathcal{H}_\mathrm{sp} $, we temporarily turn on a MW field coupling the $ \ket{\downarrow} $ and $ \ket{\uparrow} $ states (see Fig.~\ref{fig:setup}) in such a way that the electronic state is fully transferred to the lower PES eigenstate $ \ket{\varphi_-(\bm{q})}$. After that, the MW field is turned off. The subsequent dynamics in the subspace $ \mathcal{H}_\mathrm{sp} $ is then governed by $ H_\mathrm{rel}=-\nabla^2_{\bm{q}}/(2\mu)\otimes S_0+H_\mathrm{spin} $ [see Eqs.~\eqref{eq:HspinSGW} and \eqref{eq:SGW1}], with the initial state given by $ \ket{\psi_0(\bm{q})} = \ket{\psi(\bm{q},t=0)}= \phi^\mathrm{ss}_\mathrm{rel}(\bm{q})\ket{\varphi_-(\bm{q})} $. 

We now inspect the time evolution of the populations of the two PESs, $ \tilde{n}_\mu(t)$, and of the diabatic basis states, $ n_{k}(t) $. These can be obtained by expanding the full time dependent wavefunction as
\begin{equation}\label{key}
\ket{\psi(\bm{q},t)} = \sum_{k=1,2} \phi_k(\bm{q},t)\ket{\pi_k}=\sum_{\mu=\pm}\tilde{\phi}_\mu(\bm{q},t) \ket{\varphi_{\mu}(\bm{q})},
\end{equation}
where the coefficients $ \phi_k(\bm{q},t) $ [$ \tilde{\phi}_\mu(\bm{q},t) $] represent the diabatic (adiabatic) nuclear wavefunction components. Hence, the populations of the diabatic and adiabatic states are given by $ n_{k}(t)=\int|\phi_k(\bm{q},t)|^2\ d^2\bm{q} $ and $ \tilde{n}_\mu(t)=\int|\tilde{\phi}_\mu(\bm{q},t)|^2\ d^2\bm{q} $, respectively. 

To identify the effects induced by the CI, we compare the exact dynamics of the system, which takes into account the presence of the two crossing PESs, with a BO approximation, in which the nuclear motion takes place on the lower PES only. In both cases, the nuclear dynamics is obtained by solving numerically the time-dependent Schr\"odinger equation associated with $ H_\mathrm{rel} $ in the diabatic representation via a Crank-Nicholson scheme~\cite{Thomas:1995, Note1}. 

\begin{figure*}[t]
	\centering
	\includegraphics[width=\textwidth]{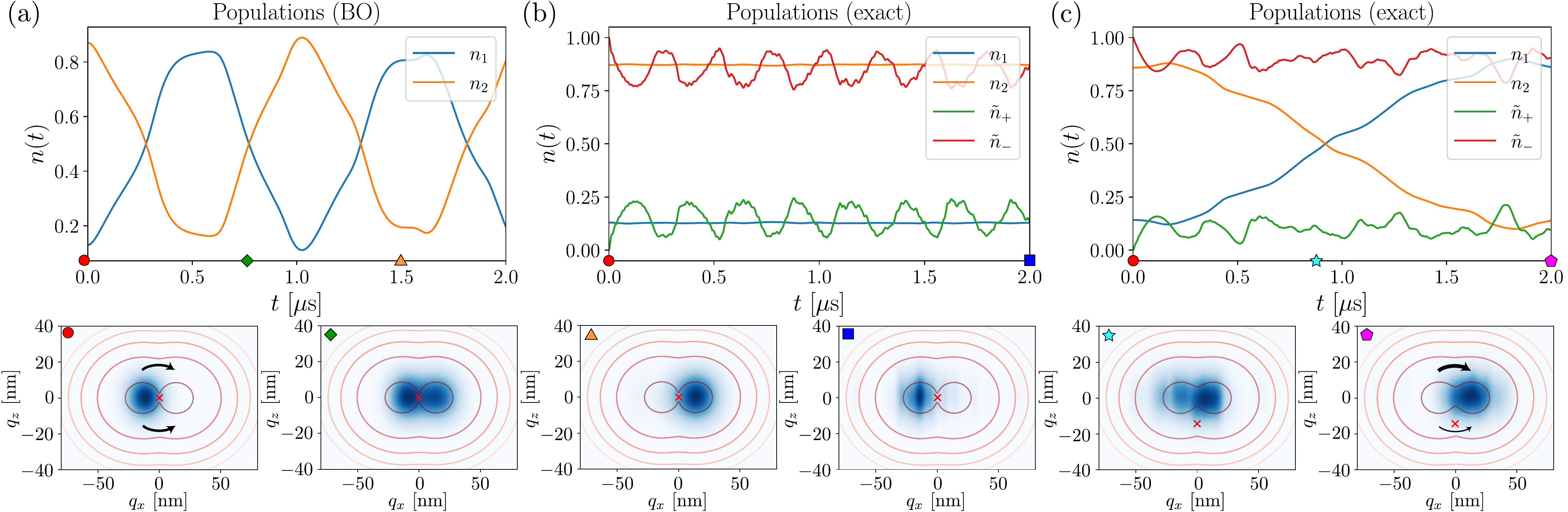}
	\caption{{\bf Dynamics of the diabatic and adiabatic populations.} Time evolution of the populations $ n_k(t) $ of the diabatic states $ \ket{\pi_1} $ (blue curve) and $ \ket{\pi_2} $ (yellow curve) in the (a) BO approximation, (b) exact dynamics with $ V^0_\mathrm{ex}/h=0 $, and (c) exact dynamics with $ V^0_\mathrm{ex}/h=2\pi\times0.3 $ MHz. In (b, c), the populations of the adiabatic states, $ \tilde{n}_-(t) $ (red curves) and $ \tilde{n}_+(t) $ (green curves) are also shown. (b): The complete destructive interference between the two paths surrounding the CI results in the freezing of the nuclear motion in the CM reference frame. (c): A finite $ V_\mathrm{ex}^0 $ shifts the CI and breaks the perfect symmetry between the two paths observed in (b). As a consequence, the nuclear wavepacket can move to the other local minimum of $ U_-(\bm{q}) $ and the diabatic populations $ n_{k}(t) $ swap with each other. The bottom row illustrates some snapshots of the dynamics of the nuclear density $ \mathcal{N}(\bm{q},t) $ in the CM reference frame associated with the various time evolutions shown in the upper row (see the corresponding marks). Here, red contours display the behavior of $ U_-(\bm{q}) $, with lighter tones corresponding to larger values, while the cross highlights the position of the CI. The arrows in the first and last plots are a sketch of the weight of the nuclear wavepackets encircling the opposite sides of the CI. In all panels, the system is initialized in the state $ \ket{\psi_0(\bm{q})} $, with $ q^0_{x,\mathrm{ss}}=-0.011\ \mu\mathrm{m} $. Other parameters are as in Fig.~\ref{fig:PES}.
	\label{fig:populations}}
\end{figure*}

In the BO approximation, the nuclei oscillate around the equilibrium position $ \bm{r}^0 $ of the relative motion and in the CM reference frame the nuclear density $ \mathcal{N}(\bm{q},t)=\sum_{k}|\phi_k(\bm{q},t)|^2 $ moves from one minimum of $ U_-(\bm{q}) $ to the other. The effects of this motion can be directly observed in the time evolution of the diabatic populations $ n_{k}(t) $, which can be monitored in real-time via a spectroscopic measurement of the Rydberg states. As shown in Fig.~\ref{fig:populations}(a), the oscillatory motion of the nuclear density gives rise to large oscillations in $ n_k(t) $. This changes drastically when the exact dynamics in the presence of the CI at $ \bm{q}^*=\bm{0} $ is considered. Due to the symmetry of the PESs, in moving from one minimum of $ U_-(\bm{q}) $ to the other, the nuclear wavepacket splits evenly on the two paths encircling the CI on the opposite sides. The relative GP accumulated in this motion around the CI on the two paths differs by $ \pi $ and results in a perfect destructive interference at the other minimum of $ U_-(\bm{q}) $. As a consequence, the nuclei are inhibited from moving: the nuclear density $ \mathcal{N}(\bm{q},t) $ is stuck in the initial minimum of $ U_-(\bm{q}) $. This has direct impact on the diabatic populations, shown in Fig.~\ref{fig:populations}(b), which exhibit only small oscillations around their initial values.

To confirm that the observed behavior is entirely due to the GP, we have also considered a BO approximation in which the diagonal BO correction terms have also been included. The latter take into account the additional potential energy barrier induced by the presence of the CI~\cite{ Gherib:2016, Ryabinkin:2017}. As shown in the SM~\cite{Note1}, the inclusion of the diagonal BO correction does not prevent the oscillatory dynamics of the nuclei (and of diabatic populations) but only results in a slight increase of the period of the oscillations. 

\textit{Controlling the CI.---} So far we considered the case in which the CI is located at the saddle point of $ U_-(\bm{q}) $ and the PESs are highly symmetric around the CI. This was achieved with a fine-tuned exchange interaction potential, such that $ V_\mathrm{ex}^0=0 $. We now examine the more general case with $ V_\mathrm{ex}^0\neq0 $. This leads to a scenario common to many real molecules, in which CIs are located on the slope of (asymmetric) PESs~\cite{Levine:2007,Yang:2018,Chang:2020}. As shown in Eq.~\eqref{eq:WQ0}, a finite value of $ V^0_\mathrm{ex} $ allows one to control the coordinate $ q^*_z $ of the CI. In this case, the symmetry of the PESs around the CI is broken and the paths encircling the two sides of the CI are no longer equivalent [Fig.~\ref{fig:PES}(b)]. This provides a clear demonstration of the fact that the freezing of the nuclear motion is due to a perfect destructive interference caused by a GP. Indeed, when $ V^0_\mathrm{ex}\neq0 $, the initial nuclear wavefunction splits in two wavepackets with different weights and the interference between the two paths encircling the CI occurs only partially. As a consequence, the nuclear density $ \mathcal{N}(\bm{q},t) $ can now move from one minimum of $ U_-(\bm{q}) $ to the other and the time evolution of the diabatic populations differs significantly from the one of Fig.~\ref{fig:populations}(b): as shown in Fig.~\ref{fig:populations}(c), $ n_1(t) $ and $ n_2(t) $ oscillate and swap with each other after a time interval proportional to $ V^0_\mathrm{ex} $.

More complex scenarios can be obtained by considering systems with more than two ions. For instance, in the SM we show that our 
approach can be readily generalized to a three-ion setup~\cite{Note1}. This provides a flexible experimental platform to study the full quantum dynamics of CIs involving more than two PESs, whose numerical investigation would require a huge amount of computational resources. 
 
\textit{Conclusions.---}
We have demonstrated that a minimal setup of two trapped Rydberg ions can be exploited to engineer a CI in a highly tunable system. An external static electric field and a tailored exchange interaction potential between Rydberg states fully control the shape of the two crossing PESs and the location of the CI. The high degree of control on the electronic and vibrational states of the ions makes it possible to simulate the dynamics of a wavepacket around the CI. We have shown that interference effects due to the GP near a CI inhibit the movement of nuclear wavepackets. This phenomenon has a direct impact on the populations of the Rydberg states and can thus be readily observed through a spectroscopic measurement. Extending this study to many-ion systems would make it possible to analyze the dynamics near CIs in more complex environments, such as in the presence of dissipation induced by coupling to many phonon modes~\cite{Kuhl:2001,Chen:2016,Ulbricht:2016}. 

\begin{acknowledgments}	
	The authors would like to thank S. W\"uster for useful discussions. The research leading to these results has received funding from the EPSRC Grant No. EP/M014266/1, the EPSRC Grant No. EP/R04340X/1 via the QuantERA project “ERyQSenS”, and the Deutsche Forschungsgemeinschaft (DFG) within the SPP 1929 Giant interactions in Rydberg Systems (GiRyd) under Project No. 428276754. CZ and MH acknowledge support from the Swedish Research Council (TRIQS), the QuantERA ERA-NET Cofund in Quantum Technologies (ERyQSenS), and the Knut \& Alice Wallenberg Foundation (WACQT). CZ acknowledges the hospitality of the University of Nottingham. IL acknowledges funding from the “Wissenschaftler R\"uckkehrprogramm GSO/CZS” of the Carl-Zeiss-Stiftung and the German Scholars  Organization e.V. WL acknowledges funding from the UKIERI-UGC Thematic Partnership No. IND/CONT/G/16-17/73, and the Royal Society through the International Exchanges Cost Share award No. IEC$\backslash$NSFC$\backslash$181078.
\end{acknowledgments}	

\footnotetext[1]{See Supplemental Material for details on the derivation of Eq.~\eqref{eq:HspinSGW}, the preparation of the nuclear initial state $ \phi^\mathrm{ss}_\mathrm{rel}(\bm{q})$, the tailored exchange interaction potential $ V_{\mathrm{ex}}(r) $, the BO approximation, and the generalization to a three-ion setup. }

\footnotetext[2]{This can be further confirmed by a self-consistency check. Using the parameters given later in the text, we obtain $  || H_\mathrm{spin}^0||/\hbar \sim 2\pi \times 10^5 $ MHz and $ || H_\mathrm{spin}^1||/\hbar \sim 2\pi  $ MHz, with $ ||\cdot|| $ being the matrix spectral norm.}

\bibliography{CI_bibliography.bib} 

\pagebreak

\widetext
\begin{center}
	\textbf{\large Supplemental Material for "Exploring the many-body dynamics near a conical intersection with trapped Rydberg ions"}
\end{center}

\setcounter{equation}{0}
\setcounter{figure}{0}
\setcounter{table}{0}
\makeatletter
\renewcommand{\theequation}{S\arabic{equation}}
\renewcommand{\thefigure}{S\arabic{figure}}

\renewcommand{\bibnumfmt}[1]{[S#1]}


\par
\begingroup
\leftskip2cm
\rightskip\leftskip
\small In this Supplemental Material we provide additional details on the derivation of Eq.~(4) of the main text, the preparation of the initial state for the system dynamics of Fig.~3 of the main text, the tailored exchange interaction potential we used in Eq.~(5) of the main text, the inclusion of the diagonal contributions to the Born-Oppenheimer approximation, and the generalization of our approach to study CIs involving multiple PESs in a three-ion setup. 
\par
\endgroup

\section{Full system dynamics}
In this section we provide additional details about the derivation of Eq.~(4) of the main text. As we state in the latter, the spin sector Hamiltonian can be written as $ H_\mathrm{spin}=H^0_\mathrm{spin}+H^1_\mathrm{spin} $, where $ H^1_\mathrm{spin} $ can be treated with a perturbative approach. The equilibrium positions of the ions can then be obtained from $ H^0_\mathrm{spin} $ and are denoted by $ \bm{R}^0=(X^0,0) $ and $ \bm{r}^0=(0,z^0) $. We then perform a small oscillation approximation around the latter. In terms of the CM and relative displacement coordinates, $ \bm{Q}=\bm{R}-\bm{R}^0 $ and $ \bm{q}=\bm{r}-\bm{r}^0 $, respectively, we obtain 
\begin{equation}\label{SM:eq:H0}
H^0_\mathrm{spin} \approx S(\bm{Q},\bm{q})\otimes S_0, \qquad \text{with} \qquad S(\bm{Q},\bm{q})=\frac{1}{2}M\bm{Q}^T\mathcal{K}^Q\bm{Q} +  \frac{1}{2}\mu \bm{q}^T \mathcal{K}^q \bm{q},
\end{equation}
where we used the vector notation $ (\bm{Q},\bm{q}) = (Q_x,Q_z,q_x,q_z) $. Here, we have introduced the dynamical matrices 
\begin{equation}\label{SM:eq:K}
(\mathcal{K}^Q)_{\alpha \beta}=\frac{1}{M}\frac{\partial^2 V_\mathrm{CM}}{\partial_{R_\alpha}\partial_{R_\beta}}\bigg|_{\bm{R}^0,\bm{r}^0} \quad \text{and} \quad (\mathcal{K}^q)_{\alpha \beta}=\frac{1}{\mu}\frac{\partial^2 V_\mathrm{rel}}{\partial_{r_\alpha}\partial_{r_\beta}}\bigg|_{\bm{R}^0,\bm{r}^0},
\end{equation}
with $ \alpha,\beta\in\{x,z\} $, and $ V_\mathrm{CM} $ and $ V_\mathrm{rel} $ given in Eq.~(3) of the main text. Explicitly, we obtain 
\begin{equation}\label{SM:eq:dynamical}
\mathcal{K}^Q=\begin{pmatrix} 
\bar{\omega}_{X}^2& 0\\
0 & \omega_Z^2
\end{pmatrix}, \quad \text{and} \quad
\mathcal{K}^q=\begin{pmatrix} 
\bar{\omega}_{X}^2 - \frac{ke^2}{\mu|z^0|^3}& 0\\
0 & \omega_Z^2  + \frac{2ke^2}{\mu|z^0|^3}
\end{pmatrix}.
\end{equation}
Finally, we write also $ H^1_\mathrm{spin} $ in terms of the displacements from the equilibrium position and expand $ V_\mathrm{ex}(|\bm{r}^0+\bm{q}|) $ to the first order in $ \bm{q} $~\footnote{Note that the second order term would result in a negligible renormalization of the quadratic contribution contained in Eq.~\eqref{SM:eq:H0}.}. The full spin Hamiltonian becomes
\begin{equation}\label{SM:eq:HspinSGW}
H_\mathrm{spin} \approx S(\bm{Q},\bm{q}) \otimes S_0 + G(\bm{Q},\bm{q}) \otimes S_z + W(\bm{q}) \otimes S_x,
\end{equation}
where we introduced
\begin{subequations}\label{SM:eq:SGW1}
	\begin{align}
	G(\bm{Q},\bm{q})&= A^2 \rho_- (Q_x+X^0) q_x,\\
	W(\bm{q})&= V_\mathrm{ex}(r^0)+\nabla_{\bm{r}}V_\mathrm{ex}(r)|_{\bm{r}=\bm{r}^0}\cdot \bm{q} \equiv V^0_\mathrm{ex}+F_z^0 q_z.\label{eq:WQ0}
	\end{align}
\end{subequations}
CIs occur at positions $ (\bm{Q}^*,\bm{q}^*) $ where the eigenvalues of $ H_\mathrm{spin} $, i.e. the potential energy surfaces (PESs), are degenerate~\cite{Ryabinkin:2017, Domcke:2004}. The latter are given by \begin{equation} U_\pm(\bm{Q},\bm{q})=S(\bm{Q},\bm{q})\pm\sqrt{G^2(\bm{Q},\bm{q})+W^2(\bm{q})} 
\end{equation}
and, hence, CIs emerge when $ G(\bm{Q}^*,\bm{q}^*)=W(\bm{q}^*)=0 $. From Eq.~\eqref{SM:eq:SGW1} we obtain that this identity is satisfied by two possibilities: $ Q^*_x = -X^0, q^*_z=-V^0_\mathrm{ex}/F_z(\bm{r}^0)  $ or $ q^*_x=0, q^*_z=-V^0_\mathrm{ex}/F_z(\bm{r}^0) $. The latter case is of particular interest since it can be studied entirely in the CM reference frame. This is seen as follows: a CI occurs in the \emph{branching plane}, which is the subspace of the nuclear coordinates where the degeneracy between the PESs is lifted linearly~\cite{Yarkony:2005}. For a generic CI located at $ (\bm{Q}^*,\bm{q}^*) $, the latter is spanned by the vectors $ \bm{u}=\bm{g}(\bm{Q}^*,\bm{q}^*)/| \bm{g}(\bm{Q}^*,\bm{q}^*) | $ and $ \bm{v}=\bm{w}(\bm{Q}^*,\bm{q}^*)/|\bm{w}(\bm{Q}^*,\bm{q}^*)| $, with 
\begin{subequations}
	\begin{align}\label{key}
	\bm{g}(\bm{Q},\bm{q})&=\nabla_{\bm{Q},\bm{q}}G(\bm{Q},\bm{q})=\alpha^2 \rho_- (q_x,0,Q_x+X^0,0),\\
	\bm{w}(\bm{Q},\bm{q})&=\nabla_{\bm{Q},\bm{q}}W(\bm{q})=F_z^0(0,0,0,1),
	\end{align}
\end{subequations}
being the gradients of $ G(\bm{Q},\bm{q}) $ and $ W(\bm{q}) $. Considering the CI located at $ \bm{q}^*=(0,-V^0_\mathrm{ex}/F_z^0) $, one finds $ \bm{u}=(0,0,1,0) $ and $ \bm{v}=(0,0,0,1) $. Thus, the corresponding branching plane coincides with the subspace of the relative nuclear coordinate and the position of the CI is not affected by the CM degrees of freedom. Moreover, by denoting with $ \ell^Q_x=(\hbar/M\mathcal{K}^Q_{xx})^{1/2} $ the characteristic oscillation length associated with the motion of the CM [see Eq.~\eqref{SM:eq:HspinSGW}], one typically finds $ X_0\gg\ell^Q_x$. Thus, along the transverse direction, the CM performs small oscillations around its equilibrium position $ X^0 $ which, in turn, results only in a little deformation of the PESs. To inspect the main effects induced by the CI at $ \bm{q}^*=(0,-V^0_\mathrm{ex}/F_z^0) $ we can therefore neglect the motion of the CM, set $ \bm{Q}=\bm{0} $ in Eq.~\eqref{SM:eq:SGW1}, and omit the dependence from $ \bm{Q} $. In this way, we obtain Eqs.~(4) and~(5) of the main text.

\section{Initialization of the nuclear wavepacket}
In this section, we discuss the initialization protocol of the nuclear wavefunction for the two-ion system for the dynamics shown in Fig.~3 of the main text. First, both the ions are adiabatically excited from their ground state $ \ket{g} $ to the $ \ket{\downarrow}=\ket{nS} $ state. Here, the full system Hamiltonian has the same form as Eq.~(1) of the main text,
\begin{equation}\label{SM:eq:Hss}
H^\mathrm{ss} = \left(-\frac{\nabla^2_{\bm{R}}}{2M}-\frac{\nabla^2_{\bm{r}}}{2\mu}\right) \otimes S_0 + H^\mathrm{ss}_\mathrm{spin},
\end{equation}
but without the exchange interaction and the coupling between the centre of mass (CM) and relative coordinates. In particular, 
\begin{equation}\label{SM:eq:Hspinss}
H^\mathrm{ss}_\mathrm{spin}=\left(V_\mathrm{CM}+V_\mathrm{rel}\right)\otimes S_0,
\end{equation}
with
\begin{subequations}\label{SM:eq:VCMVrel}
	\begin{align}
	V^\mathrm{ss}_\mathrm{CM}&=\frac{M}{2}\left[\bar{\omega}_{\mathrm{ss},X}^2\left(X-X^0_\mathrm{ss}\right)^2+\omega_Z^2 Z^2\right], \\
	V^\mathrm{ss}_\mathrm{rel}&=\frac{\mu}{2}\left(\bar{\omega}_{\mathrm{ss},X}^2 x^2+\omega_Z^2\ z^2\right)+\frac{ke^2}{r},
	\end{align}
\end{subequations}
with $ \bar{\omega}_{\mathrm{ss},X}^2=\omega_X^2-2\rho_sA^2/m $ and $ X^0_\mathrm{ss}=-e\mathcal{E}/(m\bar{\omega}_{\mathrm{ss},X}^2) $.
The CM equilibrium position is thus given by $ \bm{R}^0_\mathrm{ss}=(X^0_\mathrm{ss},0) $, where $ X^0_\mathrm{ss} $ is controlled by the external electric field $ \bm{\mathcal{E}} $. On the other hand, the relative equilibrium position is determined by the solution of the equation $ \nabla_{\bm{r}}V^\mathrm{ss}_\mathrm{rel}|_{\bm{r}^0_\mathrm{ss}}=0 $. By symmetry consideration one obtains $ \bm{r}^0_\mathrm{ss}=(0,z^0_\mathrm{ss}) $, where $ z^0_\mathrm{ss} $ is determined by the interplay between the Coulomb repulsion and the trapping potential. Expanding Eq.~\eqref{SM:eq:VCMVrel} around ions' equilibrium positions and defining the CM and relative displacements from the latter as $ \bm{Q}_\mathrm{ss}=\bm{R}_\mathrm{ss} -\bm{R}^0_\mathrm{ss}$ and $ \bm{q}_\mathrm{ss}=\bm{r}-\bm{r}^0_\mathrm{ss} $, one gets
\begin{equation}\label{SM:eq:Vss}
V^\mathrm{ss}_\mathrm{CM}+V^\mathrm{ss}_\mathrm{rel} \approx \frac{1}{2}M \bm{Q}^T_\mathrm{ss}\mathcal{K}^Q_\mathrm{ss}\bm{Q}_\mathrm{ss} + \frac{1}{2}\mu \bm{q}^T_\mathrm{ss}\mathcal{K}^q_\mathrm{ss}\bm{q}_\mathrm{ss},
\end{equation}
where the dynamical matrices $ \mathcal{K}^Q_\mathrm{ss} $ and $ \mathcal{K}^q_\mathrm{ss} $ are given by
\begin{equation}\label{SM:eq:Kss}
(\mathcal{K}^Q_\mathrm{ss})_{\alpha \beta}=\frac{1}{M}\frac{\partial^2 V^\mathrm{ss}_\mathrm{CM}}{\partial_{R_\alpha}\partial_{R_\beta}}\bigg|_{\bm{R}^0_\mathrm{ss},\bm{r}^0_\mathrm{ss}} \quad \text{and} \quad (\mathcal{K}^q_\mathrm{ss})_{\alpha \beta}=\frac{1}{\mu}\frac{\partial^2 V^\mathrm{ss}_\mathrm{rel}}{\partial_{r_\alpha}\partial_{r_\beta}}\bigg|_{\bm{R}^0_\mathrm{ss},\bm{r}^0_\mathrm{ss}}.
\end{equation}
with $ \alpha,\beta\in\{x,z\} $. In particular, from Eq.~\eqref{SM:eq:VCMVrel} we obtain 
\begin{equation}\label{SM:eq:dynamicalSS}
\mathcal{K}^Q_\mathrm{ss}=\begin{pmatrix} 
\bar{\omega}_{\mathrm{ss},X}^2& 0\\
0 & \omega_Z^2
\end{pmatrix},\quad \text{and} \quad
\mathcal{K}^q_\mathrm{ss}=\begin{pmatrix} 
\bar{\omega}_{\mathrm{ss},X}^2 - \frac{ke^2}{\mu|z^0_\mathrm{ss}|^3}& 0\\
0 & \omega_Z^2  + \frac{2ke^2}{\mu|z^0_\mathrm{ss}|^3}
\end{pmatrix}.
\end{equation}
Therefore, being $ \mathcal{K}^Q_\mathrm{ss} $ and $ \mathcal{K}^q_\mathrm{ss} $ diagonal, Eq.~\eqref{SM:eq:Vss} can be written as a sum of harmonic potentials with renormalized trapping frequencies,
\begin{equation}\label{key}
V_\mathrm{CM}+V_\mathrm{rel}\approx\frac{1}{2}M\left(\bar{\omega}_{\mathrm{ss},X}^2 Q_{\mathrm{ss},x}^2 + \omega_Z^2 Q_{\mathrm{ss},z}^2 \right)+\frac{1}{2}\mu\left(\tilde{\omega}_X^2 q_{\mathrm{ss},x}^2 + \tilde{\omega}_Z^2 q_{\mathrm{ss},z}^2 \right),
\end{equation}
where
\begin{equation}\label{key}
\tilde{\omega}_{\mathrm{ss},X}= \sqrt{\bar{\omega}_{\mathrm{ss},X}^2-\frac{ke^2}{\mu|z^0_\mathrm{ss}|^3}}, \quad\text{and}\quad \tilde{\omega}_{\mathrm{ss},X}= \sqrt{\omega_Z^2+\frac{2ke^2}{\mu|z^0_\mathrm{ss}|^3}}.
\end{equation}
The (nuclear) ground-state wavefunction associated with this potential is thus given by
\begin{equation}\label{SM:eq:ssGS}
\Phi^\mathrm{ss}(\bm{Q},\bm{q}) = \phi^\mathrm{ss}_\mathrm{CM}(\bm{Q})\phi^\mathrm{ss}_\mathrm{rel}(\bm{q}),
\end{equation}
with
\begin{align}\label{SM:eq:ssstate}
\phi^\mathrm{ss}_\mathrm{CM}(\bm{Q}) = \frac{\exp\left[-\frac{1}{2}\left(\frac{Q_x}{\ell_{\mathrm{ss},x}^Q}\right)^2-\frac{1}{2}\left(\frac{Q_z}{\ell_{\mathrm{ss},z}^Q}\right)^2\right]}{\sqrt{\pi \ell_{\mathrm{ss},x}^Q \ell_{\mathrm{ss},z}^Q}}  \quad \text{and} \quad
\phi^\mathrm{ss}_\mathrm{rel}(\bm{q}) = \frac{\exp\left[-\frac{1}{2}\left(\frac{q_x-q^0_x}{\ell_{\mathrm{ss},x}^q}\right)^2-\frac{1}{2}\left(\frac{q_z}{\ell_{\mathrm{ss},z}^q}\right)^2\right]}{\sqrt{\pi \ell_{\mathrm{ss},x}^q\ell_{\mathrm{ss},z}^q}},
\end{align}
where $ \ell_{\mathrm{ss},\alpha}^Q= (\hbar/M \bar{\omega}_\alpha)^{1/2} $ and $ \ell_{\mathrm{ss},\alpha}^q= (\hbar/\mu \tilde{\omega}_\alpha)^{1/2} $ are the typical harmonic oscillator lengthscales. In Eq.~\eqref{SM:eq:ssstate}, we have also included the displacements $ q^0_{\mathrm{ss},x} $, which is crucial for the investigation of the effects induced by the conical intersection (CI) in the main text. The latter can be achieved, e.g., by exciting the transverse breathing mode of the two-ion system or by modifying the electric field $ \bm{\mathcal{E}}=(\mathcal{E},0) $ to $ \bm{\mathcal{E}}=(\mathcal{E}+Cx,Cz) $, with $ C $ being a constant. In the latter case, it is possible to show that in order to obtain a finite $ q^0_{\mathrm{ss},x} $ of the order of few nanometers the corrections to the analysis above [and, in particular, to Eq.~\eqref{SM:eq:dynamicalSS}] are negligible. \\

We conclude this section by discussing the relationship between the initial Gaussian nuclear wavefunction and the symmetry properties of the system. We recall that, in the CM reference frame and neglecting the motion of the CM, the dynamics of the system is governed by $ H_\mathrm{rel}=-\nabla^2_{\bm{q}}/(2\mu)\otimes S_0 +H_\mathrm{spin} $, with $ H_\mathrm{spin} $ given in Eq.~(1) of the main text. In general, the eigenstates of $ H_\mathrm{rel} $ are given by
\begin{equation}\label{key}
H_\mathrm{rel}\ket{\psi^E(\bm{q})}=E\ket{\psi^E(\bm{q})},
\end{equation}
with $ \ket{\psi^E(\bm{q})} $ belonging to the Hilbert space $ \mathcal{H}=\mathcal{H}_{\bm{q}}\otimes\mathcal{H}_\mathrm{spin} $, where $\mathcal{H}_{\bm{q}}$ is the Hilbert space of a single particle moving in the two-dimensional CM frame and ${H}_\mathrm{spin}$ is the two-dimensional Hilbert space associated with its spin $ 1/2 $ degree of freedom.
In the adiabatic representation, the eigenstate $ \ket{\psi^E(\bm{q})} $ can be written as
\begin{equation}\label{key}
\ket{\psi^E(\bm{q})}=\sum_{\mu=\pm} \tilde{\phi}^E_\mu(\bm{q})\ket{\varphi_{\mu}(\bm{q})},
\end{equation}
with $ \ket{\varphi_{\mu}(\bm{q})} $ defined in Eq.~(8) of the main text. The functions $ \tilde{\phi}_\mu^E(\bm{q}) $ are called nuclear wavefunctions, despite one has to keep in mind that they are not proper wavefuctions on $ \mathcal{H}_{\bm{q}} $~\cite{Bohm:2003}.

From Eq.~(5) of the main text one can verify that, for $ V^0_\mathrm{ex}=0 $, the parity operator $ \mathcal{P} $ implementing the spatial reflection $ \bm{q}\rightarrow-\bm{q} $ is a symmetry of the system, i.e., $ [H_\mathrm{rel},\mathcal{P}]=0 $. Therefore, it is possible to find a basis in $ \mathcal{H} $, denoted by $ \{\ket{\chi_+^E(\bm{q})},\ket{\chi_-^E(\bm{q})}\}  $, such that
\begin{subequations}
	\begin{align}
	H_\mathrm{{rel}}\ket{\chi_\pm^E(\bm{q})}=E\ket{\chi_\pm^E(\bm{q})},\label{SM:eq:chipm}\\
	\mathcal{P}\ket{\chi_\pm^E(\bm{q})}=\pm\ket{\chi_\pm^E(\bm{q})},
	\end{align} 
\end{subequations}
where, in the adiabatic basis, 
\begin{equation}\label{SM:eq:wfchipm}
\ket{\chi^{E}_{\pm}(\bm{q})}=\sum_{\mu=\pm}\tilde{\xi}^{E}_{\mp,\mu}(\bm{q})\ket{\varphi_{\mu}(\bm{q})}.
\end{equation}
Due to the presence of the CI at $ \bm{q}=\bm{0} $ and of the associated geometric phase, one has $ \mathcal{P}\ket{\varphi_{\mu}(\bm{q})}=-\ket{\varphi_{\mu}(\bm{q})},\, \forall \bm{q} $. Therefore, to preserve the single-valuedness of $ \ket{\chi^E_\pm(\bm{q})} $, the functions $ \tilde{\xi}^{E}_{\pm,\mu}(\bm{q}) $ must obey the identity $ \mathcal{P}\tilde{\xi}^{E}_{\pm,\mu}(\bm{q})=\mp\tilde{\xi}^{E}_{\pm,\mu}(\bm{q}) $, i.e., $ \tilde{\xi}^E_{+,\mu}(\bm{q}) $ [$ \tilde{\xi}^E_{-,\mu}(\bm{q}) $] is even (odd) with respect to the transformation $ \bm{q}\rightarrow-\bm{q} $.

We can now focus on the initial state, $ \ket{\psi_{0}(\bm{q})}=\phi^{\mathrm{ss}}_{\mathrm{rel}}(\bm{q})\ket{\varphi_-(\bm{q})} $. Since $ \phi^{\mathrm{ss}}_{\mathrm{rel}}(\bm{q}) $ is a Gaussian peaked in one of the two minima of the lower PES, the initial state $ \ket{\psi_{0}(\bm{q})} $ can be expanded to a high degree of approximation on the basis $ \{\ket{\chi_+^E(\bm{q})},\ket{\chi_-^E(\bm{q})}\} $ as
\begin{equation}\label{key}
\ket{\psi_{0}(\bm{q})}\approx\sum_{E}c_E(0)\left(\ket{\chi_+^E(\bm{q})}+\ket{\chi_-^E(\bm{q})}\right),
\end{equation}
with $ c_E(0) $ a given coefficient. Note that the choice of the sign between the $ \ket{\chi_\pm^E(\bm{q})} $ states is arbitrary and the expansion with the $ - $ sign would result in the state in which the nuclear wavefunction is centered in the other minimum of $ U_-(\bm{q}) $. Indeed, the equally weighted superpositions $ \tilde{\xi}^E_{+,\mu}(\bm{q}) \pm \tilde{\xi}^E_{-,\mu}(\bm{q})$ are non-vanishing in one of the two minima of $ U_-(\bm{q}) $ only. Since $ \ket{\chi_\pm(\bm{q})} $ are eigenstates of $ H_\mathrm{rel} $, the initial wavefunction $ \ket{\psi_{0}(\bm{q})} $ evolves in time as
\begin{equation}\label{SM:eq:psi0evo}
\ket{\psi_{0}(\bm{q},t)}\approx\sum_{E}c_E(t)\left(\ket{\chi_+^E(\bm{q})}+\ket{\chi_-^E(\bm{q})}\right),
\end{equation}
with $ c_E(t)=e^{iEt}c_E(0) $. This implies that $ \ket{\psi_{0}(\bm{q},t)} $ is an equally weighted superposition of $ \ket{\chi_+^E(\bm{q})} $ and $ \ket{\chi_-^E(\bm{q})} $ $ \forall t $ and, hence, it remains in the same minimum of $ U_-(\bm{q}) $ up to very long timescales.

In the case with $ V^0_\mathrm{ex}\neq 0 $, $ \mathcal{P} $ is no longer a symmetry of the system and, in general, its eigenstates $ \ket{\chi_\pm(\bm{q})} $ are not eigenstates of $ H_\mathrm{rel} $. Thus, the time evolved initial nuclear wavepacket cannot be written in the form of Eq.~\eqref{SM:eq:psi0evo}: $ \ket{\psi_0(\bm{q},t)} $ is not an equally weighted superposition of the states $ \ket{\chi_\pm(\bm{q})} $ and, in general, it is not constrained to remain in one of the minima of $ U_-(\bm{q}) $.

\section{Microwave dressed exchange potential}
\begin{figure*}[t]
	\centering
	\includegraphics[width=\textwidth]{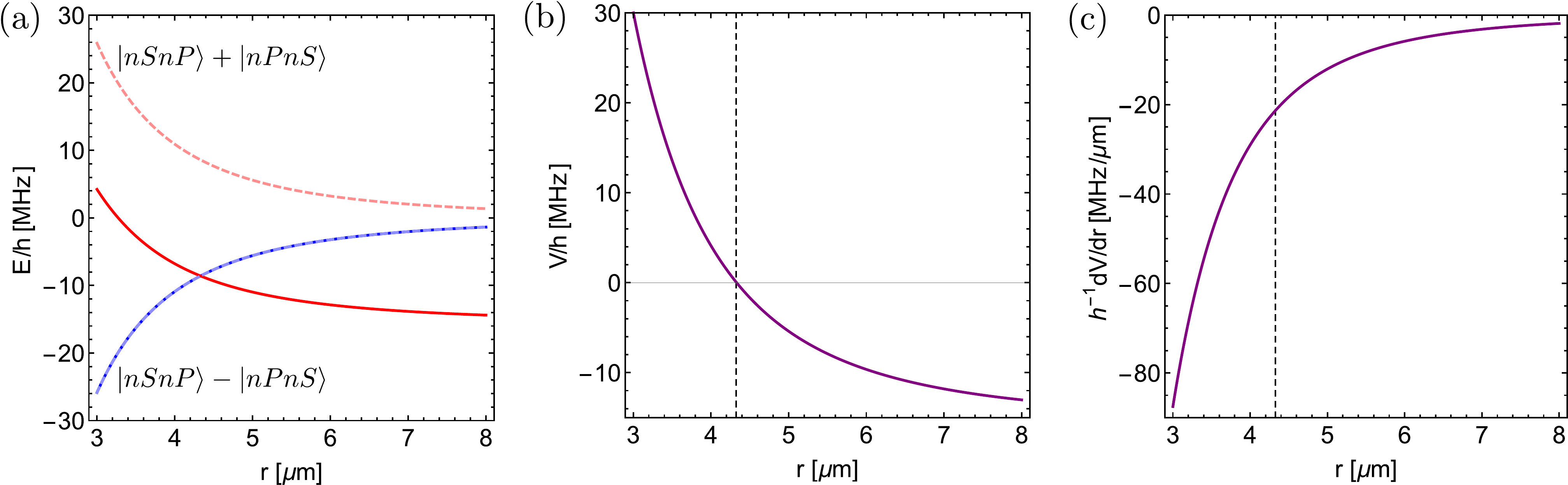}
	\caption{{\bf MW dressed potential.} (a) Eigenvalues of $ \mathcal{V}_\mathrm{ex} $, $ E_{-}(r) $ (dashed light blue) and $ E_{+} $ (dashed light red) (corresponding to the states $ \ket{-} $ and $ \ket{+} $), as a function of the ion-ion distance $ r $. A bichromatic MW field shifts the energy of the $ \ket{nPnP} $ state by $ \delta E_\mathrm{PP}=2\pi\times697.94\ \mathrm{MHz} $ (not shown). In the presence of second red-detuned MW field with Rabi frequency $ \Omega_\mathrm{MW}=2\pi\times43 \ \mathrm{MHz} $ and detuning $ \Delta_{\mathrm{MW}}=-2\pi\times50\ \mathrm{MHz} $, the energy of the bare eigenvalue $ E_{+}(r) $ is lowered (solid red), while the one of $ E_{-}(r) $ is unchanged (solid blue). (b) Effective exchange interaction potential $ V_\mathrm{ex}(r) $ as a function of the ion-ion distance $ r $, obtained by subtracting the shifted values of $ E_+(r) $ and $ E_-(r) $ given by the solid curves of panel (a). (c) Gradient $ |\nabla_{\bm{r}} V_\mathrm{ex}(r)| $ as a function of $ r $. In panels (b) and (c), the vertical dashed line corresponds to $ r=4.3\ \mathrm{\mu m}\approx r^0 $, where $ V_\mathrm{ex}(r^0)=0 $ and $ |\nabla_{\bm{r}}V_\mathrm{ex}(\bm{r})|_{\bm{r}=\bm{r}^0}\neq 0 $. In all panels, $ n=50 $ and $ C_3=697.94\ \mathrm{MHz\ \mu m^3} $. \label{SM:fig:potential}}
\end{figure*}

In this section we provide additional details about the microwave (MW) dressing scheme employed to engineer the exchange interaction potential $ V_\mathrm{ex}(r) $ we used in Eq.~(5) of the main text. We consider a pair of $ ^{88}\mathrm{Sr}^+ $ trapped Rydberg ions, and focus on their Rydberg states $ \ket{nS}=\ket{\downarrow}=(0, 1)^T $ and $ \ket{nP}=\ket{\uparrow}=(1, 0)^T $ (with $ n $ a positive integer). The latter interact through a conventional exchange interaction potential 
\begin{equation}\label{key}
\mathcal{V}_\mathrm{ex}=\tilde{V}_\mathrm{ex}(r) (\sigma_+^1\otimes\sigma_-^2+\sigma_-^1\otimes\sigma_+^2),
\end{equation}
with $ \tilde{V}_\mathrm{ex}(r)=C_3/r^3 $ and $ r=|\bm{r}| $ being the distance between the ions. The corresponding eigenstates are $ \ket{nSnS} $, $ \ket{-}= (\ket{nSnP}-\ket{nPnS})/\sqrt{2} $, $ \ket{+}=(\ket{nSnP}+\ket{nPnS})/\sqrt{2}  $, $ \ket{nPnP} $. In particular, the eigenvalues $ E_\pm(r) $ associated with $ \ket{-} $ and $ \ket{+} $ are shown in Fig.~\ref{SM:fig:potential} (dashed lines). Note that $ E_+(r)>E_-(r),\ \forall r $. 

We now consider the following MW dressing scheme. First, we add a bichromatic MW field coupling the state $ \ket{nP} $ to another state $ \ket{n'S} $, with $ n'\neq n $, by a two-photon transition. The two frequencies of the bichromatic field are chosen so that they have opposite and symmetric detuning with respect to the $ \ket{nP} $ to $ \ket{n'S} $ transition. Thus, the bichromatic MW field causes no shift to the $ \ket{nP} $ level of a single ion. However, in the presence of interactions between the $ \ket{nP} $ and $ \ket{n'S} $ states of a two-ion system, the pair state $ \ket{nPn'S}+\ket{n'SnP} $ is shifted. As a consequence, the bichromatic MW field is no longer symmetrically detuned with respect to the transition from $ \ket{nPnP} $ to $ \ket{nPn'S}+\ket{n'SnP} $ and causes a shift of these two levels. In our two-state description, this energy shift can be effectively modeled by adding a contribution $ \delta\mathcal{V}_{\mathrm{PP}}=\delta E_\mathrm{PP} \ket{nPnP}\bra{nPnP} $ to the two-ion Hamiltonian. 

Next, we add another red-detuned MW field coupling single-ion $ \ket{nS} $ and $ \ket{nP} $ states with Rabi frequency $ \Omega_{\mathrm{MW}} $ and detuning $ \Delta_{\mathrm{MW}} $. The single-ion Hamiltonian is 
\begin{equation}\label{key}
H_\mathrm{MW}=\frac{1}{2}\begin{pmatrix}
\Delta_{\mathrm{MW}}  & \Omega_{\mathrm{MW}} \\
\Omega_{\mathrm{MW}} & - \Delta_{\mathrm{MW}}
\end{pmatrix}.
\end{equation}
The overall Hamiltonian for the two-ion system thus reads
\begin{equation}\label{SM:eq:MW}
H_\mathrm{ex}=H^{(1)}_\mathrm{MW}\otimes\mathbb{I}_2+\mathbb{I}_2\otimes H^{(2)}_\mathrm{MW}+ \mathcal{V}_\mathrm{ex}+\delta\mathcal{V}_{\mathrm{PP}},
\end{equation}
with $ \mathbb{I}_2 $ the $ 2\times2 $ identity matrix. The red-detuned MW field couples the $ \ket{nSnS} $ and $ \ket{+} $ states. Since $ \ket{nPnP} $ is far detuned (due to the presence of the bichromatic MW field), this coupling results in a shift of the $ \ket{+} $ state. On the contrary, $ \ket{-} $ is not coupled to any state by the red-detuned MW field and does not experience any shift. In particular, when $ \Delta<0 $, the energy of $ \ket{+} $ is lowered in a such a way that $ E_+(r) $ and $ E_-(r) $ cross at a given ion-ion distance $ r^* $ [see solid curves in Fig.~\ref{SM:fig:potential}(a)]. The effective two-ion exchange potential $ V_\mathrm{ex}(r) $ is readily obtained by subtracting $ E_+(r) $ and $ E_-(r) $ and it is shown in Fig.~\ref{SM:fig:potential}(b). Clearly, $ V_\mathrm{ex}(r^*)=0 $, with $ r^* $ denoted by the vertical dashed line in panels (b) and (c). By properly tuning $ \Omega_{\mathrm{MW}} $ and $ \Delta_{\mathrm{MW}} $ it is possible to make $ r^* $ coinciding with the distance $ r^0 $ between the ions in their equilibrium positions (see main text). In particular, from Fig.~\ref{SM:fig:potential}(c), one can see that this dressing scheme allows us to obtain a gradient $ h^{-1}|F_z^0|=h^{-1}|\nabla_{\bm{r}}V_\mathrm{ex}(r)|_{\bm{r}=\bm{r}^0} $ of the order of $ 2\pi\times20\ \mathrm{MHz/\mu m} $, as the one we employed in the main text. \\

Note that the potential $ V_\mathrm{ex}(r) $ shown in Fig.~\ref{SM:fig:potential}(b) does not vanish at infinite distance, as it should. The reason of this fact is that, in our model, we assumed that the effect of the bichromatic MW field is to induce a large shift in the energy of the $ \ket{nPnP} $ state. As a consequence, since $ \ket{nPnP} $ is out of resonance, the second red-detuned MW field of Eq.~\eqref{SM:eq:MW} only couples $ \ket{nSnS} $ to $ \ket{+} $, allowing us to realize the crossing between $ E_+(r) $ and $ E_-(r) $ shown in Fig.~\ref{SM:fig:potential}(a). However, the assumption that the $ \ket{nPnP} $ level is shifted away from the other states by a significant amount of energy relies on the fact that the bichromatic MW coupling strength between $ \ket{n'Sn'S} $ and $ \ket{nPnP} $ is not zero which, in turn, requires a finite interaction between the $ \ket{n'S} $ and $ \ket{nP} $ Rydberg states. This condition can only be satisfied at finite distances. Therefore, in the long distance limit, the shift of the $ \ket{nPnP} $ state (and, consequently, the one of $ \ket{+} $) will vanish in such a way that $ V_\mathrm{ex}(r)\rightarrow0 $ for $ r\rightarrow\infty $. 

\section{Born-Oppenheimer approximation with the diagonal correction}
\begin{figure*}[t]
	\centering
	\includegraphics[width=0.8\textwidth]{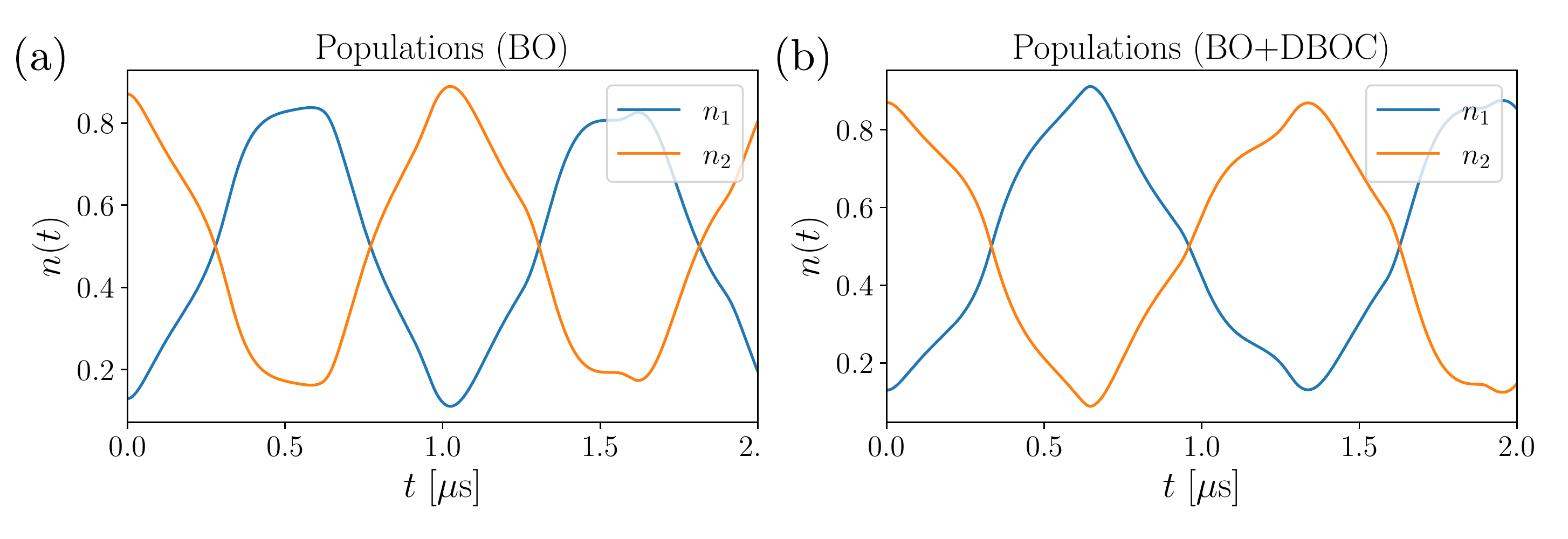}
	\caption{{\bf Dynamics of the diabatic population in the BO and BO+DBOC approximation.} Time evolution of the populations $ n_k(t) $ of the diabatic states $ \ket{\pi_1} $ (blue curve) and $ \ket{\pi_2} $ (yellow curve) in the (a) BO approximation [same as Fig.~3(a) of the main text] and (b) BO+DBOC approximation with $ V^0_\mathrm{ex}=0 $. (b) Despite the presence of the addition potential energy contribution due to the DBOC, the diabatic populations still oscillate and swap with each other. The only difference with respect to the BO approximation of panel (a) is that the period of the oscillations is slightly increased. The system is initialized on the lower PES in the state $ \ket{\psi_0(\bm{q})} $ (see main text), with $ X^0_\mathrm{ss}=-0.024\ \mu\mathrm{m} $, $ Z^0_\mathrm{ss}=0\ \mu\mathrm{m} $, $ x^0_\mathrm{ss}=-0.011\ \mu\mathrm{m} $, and $ z^0_\mathrm{ss}=4.31\ \mu\mathrm{m} $. Other parameters are as in Fig.~(3) of the main text. \label{SM:fig:DBOC}}
\end{figure*}

Finally, in this section we briefly comment about the Born-Oppenheimer (BO) approximation we used in the main text and, in particular, on the inclusion of the diagonal BO correction (DBOC). First, we write the total wavefunction of the system in the adiabatic representation~\cite{Wuster:2018,Ryabinkin:2017},
\begin{equation}\label{SM:eq:Psiadiabatic}
\ket{\Psi(\bm{q},t)}=\sum_{\mu=\pm}\tilde{\phi}_\mu(\bm{q},t)\ket{\varphi_\mu(\bm{q})},
\end{equation}
where $ \ket{\varphi_\mu(\bm{q})} $ are the eigenstates of the spin Hamiltonian $ H_\mathrm{spin} $ and are given in Eq.~(8) of the main text, while $ \tilde{\phi}_\mu(\bm{q},t) $ are the nuclear wavefunction components. The time evolution of $  \ket{\Psi(\bm{q},t)} $ is governed by the Schr\"odinger equation $ i\partial_t \ket{\Psi(\bm{q},t)}= H \ket{\Psi(\bm{q},t)} $, with $ H $ the full system Hamiltonian defined in Eq.~(1) of the main text. Substituting in the latter the adiabatic representation given in Eq.~\eqref{SM:eq:Psiadiabatic}, we obtain the following coupled equations governing the evolution of the adiabatic components of the nuclear wavefunction $ \tilde{\phi}_\mu(\bm{q},t) $ \cite{Wuster:2011},
\begin{equation}\label{SM:eq:adiabaticSch}
i\partial_t \tilde{\phi}_{\mu}(\bm{q},t)=\left[-\frac{\nabla_{\bm{q}}}{m}+V_\mu(\bm{q})\right]\tilde{\phi}_{\mu}(\bm{q},t)+\sum_{\nu}D_{\mu\nu}(\bm{q})\tilde{\phi}_{\nu}(\bm{q},t),
\end{equation}
where the nonadiabatic couplings are defined as 
\begin{equation}\label{key}
D_{\mu\nu}(\bm{q})=-\frac{1}{m}\left[\average{\varphi_\mu(\bm{q})}{\nabla^2_{\bm{q}}}{\varphi_\nu(\bm{q})}+2\average{\varphi_\mu(\bm{q})}{\nabla_{\bm{q}}}{\varphi_\nu(\bm{q})}\cdot\nabla_{\bm{q}}\right].
\end{equation}
In the BO approximation the last term in Eq.~\eqref{SM:eq:adiabaticSch} is neglected and the dynamics of the nuclear wavefunctions on the two potential energy surfaces (PESs) are decoupled. Focusing on the case with $ V^0_\mathrm{ex}=0 $, the DBOCs are given by the diagonal elements of the nonadiabatic coupling matrix~\cite{Gherib:2016,Ryabinkin:2017}, 
\begin{equation}\label{key}
D_{\mu\mu}(\bm{q})=-\frac{1}{m}\average{\varphi_\mu(\bm{q})}{\nabla^2_{\bm{q}}}{\varphi_\mu(\bm{q})}=-[\nabla_{\bm{q}}\Lambda(\bm{q})]^2=\frac{1}{4 m}\frac{q_x^2+q_z^2}{(\gamma q_x^2+\gamma^{-1}q_z^2)^2},
\end{equation}
where we used Eq.~(8) of the main text and defined $ \gamma=\alpha^2\rho_-X^0/F_z^0 $. In the presence of the DBOC, the motions of the two adiabatic components of the nuclear wavefunction $ \tilde{\phi}_\mu(\bm{q},t) $ are still decoupled but now they take place on the modified PESs $ \tilde{V}_\mu(\bm{q})= V_\mu(\bm{q})+D_{\mu\mu}(\bm{q})$. Note that $ D_{\mu\mu}(\bm{q})\rightarrow \infty $ for $ \bm{q}\rightarrow\bm{q}^*=\bm{0} $: the DBOC introduces an additional effective repulsive potential that takes into account the divergence of the nonadiabatic couplings at the CI. It is well-established that the DBOC overestimates the potential energy contribution due to the CI which, in general, is compensated by geometric phase (GP) effects~\cite{Gherib:2016,Ryabinkin:2017}. Thus, studying the system dynamics in the BO+DBOC approximation represents a “worst” case scenario and it allows us to assess whether the localization effect we observed in the system exact dynamics is due to the additional repulsive potential arising at the CI or to purely GP effects. By comparing the time evolution of the diabatic populations $ n_k(t) $ in the BO and BO+DBOC approximations in Fig.~\ref{SM:fig:DBOC}, one can notice that the only effect of the additional potential barrier introduced by the DBOC is to increase slightly the period of the oscillations. This, indeed, confirms that the localization of the nuclear wavepacket is entirely due to the GP. 

\section{Conical intersections in a three-ion setup}
In this section we show how the approach we developed in the main text can be generalized to setups with larger number of ions, focusing in particular on the three-ion case. Here, the potential energy of the ions in the harmonic trap becomes
\begin{equation}\label{key}
V_\mathrm{trap}=\frac{1}{2}m\left[\omega_X^2\left(X_1^2+X_2^2+X_3^2\right)+\omega_Z^2\left(Z_1^2+Z_2^2+Z_3^2\right)\right]+\frac{ke^2}{R_{12}}+\frac{ke^2}{R_{13}}+\frac{ke^2}{R_{23}},
\end{equation}
where $ \bm{R}_i=(X_i,Z_i) $, $ i\in\{1,2,3\} $ are the nuclear coordinate in the laboratory reference frame and $ R_{ij}=|\bm{R}_i-\bm{R}_j| $. Similarly, the potential energy contributions due to Rydberg states polarization and to the external electric field are $ \delta V_{\mathrm{trap},\sigma}=-\rho_\sigma A^2(X_1^2+X_2^2+X_3^2) $ and $ V_\mathrm{mm}=e\mathcal{E}(X_1+X_2+X_3) $, respectively. The single $ p- $excitation subspace $ \mathcal{H}_\mathrm{sp} $ is spanned by the electronic states $ \ket{\pi_1}=\ket{\uparrow\downarrow\downarrow} $,  $ \ket{\pi_2}=\ket{\downarrow\uparrow\downarrow} $, and $ \ket{\pi_3}=\ket{\downarrow\downarrow\uparrow} $. In this case, the Pauli matrices can be conveniently replace by the Gell-Mann matrices
\begin{equation}\label{key}
\lambda_0=\begin{pmatrix}
1 & 0 & 0 \\ 0& 1& 0\\ 0& 0& 1
\end{pmatrix}, \
\lambda_1=\begin{pmatrix}
0 & 1 & 0 \\ 1& 0& 0\\ 0& 0& 0
\end{pmatrix}, \
\lambda_3=\begin{pmatrix}
1 & 0 & 0 \\ 0& -1& 0\\ 0& 0& 0
\end{pmatrix}, \
\lambda_4=\begin{pmatrix}
0 & 0 & 1 \\ 0& 0& 0\\ 1& 0& 0
\end{pmatrix}, \
\lambda_6=\begin{pmatrix}
0 & 0 & 0 \\ 0& 0& 1\\ 0& 1& 0
\end{pmatrix}, \
\lambda_8=\frac{1}{\sqrt{3}}\begin{pmatrix}
1& 0 & 0 \\ 0& 1& 0\\ 0& 0& -2
\end{pmatrix}, 
\end{equation}
where we omitted $ \lambda_2, \lambda_5, \lambda_7 $ since they will not be used in the following.

In the laboratory reference frame, the Hamiltonian of the system is given by 
\begin{equation}\label{key}
H=\left(-\frac{\nabla_{\bm{R}_1}^2}{2m}-\frac{\nabla_{\bm{R}_2}^2}{2m}-\frac{\nabla_{\bm{R}_3}^2}{2m}\right)\otimes\lambda_0+H_\mathrm{spin},
\end{equation}
with $ H_\mathrm{spin}=H_\mathrm{spin}^0 + H_\mathrm{spin}^1 $. Here, 
\begin{align}
H_\mathrm{spin}^0&=\left\{\frac{1}{2}m\sum_{i=1}^{3}
\left[\bar{\omega}_X^2\left(X_i-X^0\right)^2+\omega_Z^2Z_i^2\right]+\sum_{i=1}^{2}\sum_{j>i}\frac{ke^2}{R_{ij}}\right\}\otimes\lambda_0\label{SM:eq:3:Hspin0}\\
H_\mathrm{spin}^1&=V_\mathrm{ex}(R_{12})\otimes\lambda_1+V_\mathrm{ex}(R_{13})\otimes\lambda_4+V_\mathrm{ex}(R_{23})\otimes\lambda_6-A^2\rho_1\left[\left(X_1^2-X_2^2\right)\otimes \lambda_3+\frac{1}{\sqrt{3}}\left(X_1^2+X_2^2-2X_3^2\right)\otimes \lambda_8\right],
\end{align}
with $ \bar{\omega}_X^2=\omega_X^2 - 2 A^2 \rho_0/m  $ and $ X^0=-2e\mathcal{E}/(m\bar{\omega}_X^2) $. Here, we introduced the weighted polarizabilities $ \rho_0=(\rho_\uparrow+2\rho_\downarrow)/3 $ and $ \rho_1=(\rho_\uparrow-\rho_\downarrow)/2 $.

As in the two-ion case, the Coulomb interaction contained in Eq.~\eqref{SM:eq:3:Hspin0} represents the largest energy scale and system. Therefore, $ H_\mathrm{spin}^1 $ can be treated as a small perturbation with respect to $ H_\mathrm{spin}^0 $. The equilibrium positions of the ions $ \bm{R}_i^0 $ are thus be determined by the equations $ \nabla_{\bm{R}_i} H_\mathrm{spin}^{0}=0,\ \forall i $ with a high degree of accuracy. Using the parameters we employed in the main text, we obtain $ \bm{R}_1^0=(X^0,-Z^0) $, $ \bm{R}_2^0=(X^0, 0) $, and $ \bm{R}_3^0=(X_0,Z_0) $. Using these values, the MW exchange potential is tailored in such a way $ V_\mathrm{ex}(R_{12}^0)=V_\mathrm{ex}(R_{23}^0)=0 $, where $ R_{ij}^0=|\bm{R}_i^0-\bm{R}_j^0| $.

Importantly, to investigate the possible emergence of CIs, we notice that $ H^0_\mathrm{spin}\propto \lambda_0 $. Its contribution to the eigenvalues of $ H_\mathrm{spin} $ is degenerate and, hence, it affects neither the presence nor the position of CIs. To address the latter, we can then focus on $ H_\mathrm{spin}^1 $ only. Furthermore, since the configuration space of the nuclei is six dimensional, we restrict our analysis to the $ \bar{X}-\bar{Z} $ plane, where the coordinates $ \bar{X} $ and $ \bar{Z} $ are defined according to $ \bm{R}_1=(X^0-\bar{X}, -\bar{Z}) $, $ \bm{R}_2=(X_0+2\bar{X},0) $, and $ \bm{R}_3=(X^0-\bar{X}, \bar{Z}) $. This corresponds to study the three-ion system along its breathing transverse and longitudinal vibrational modes.

\begin{figure*}[t]
	\centering
	\includegraphics[width=0.9\textwidth]{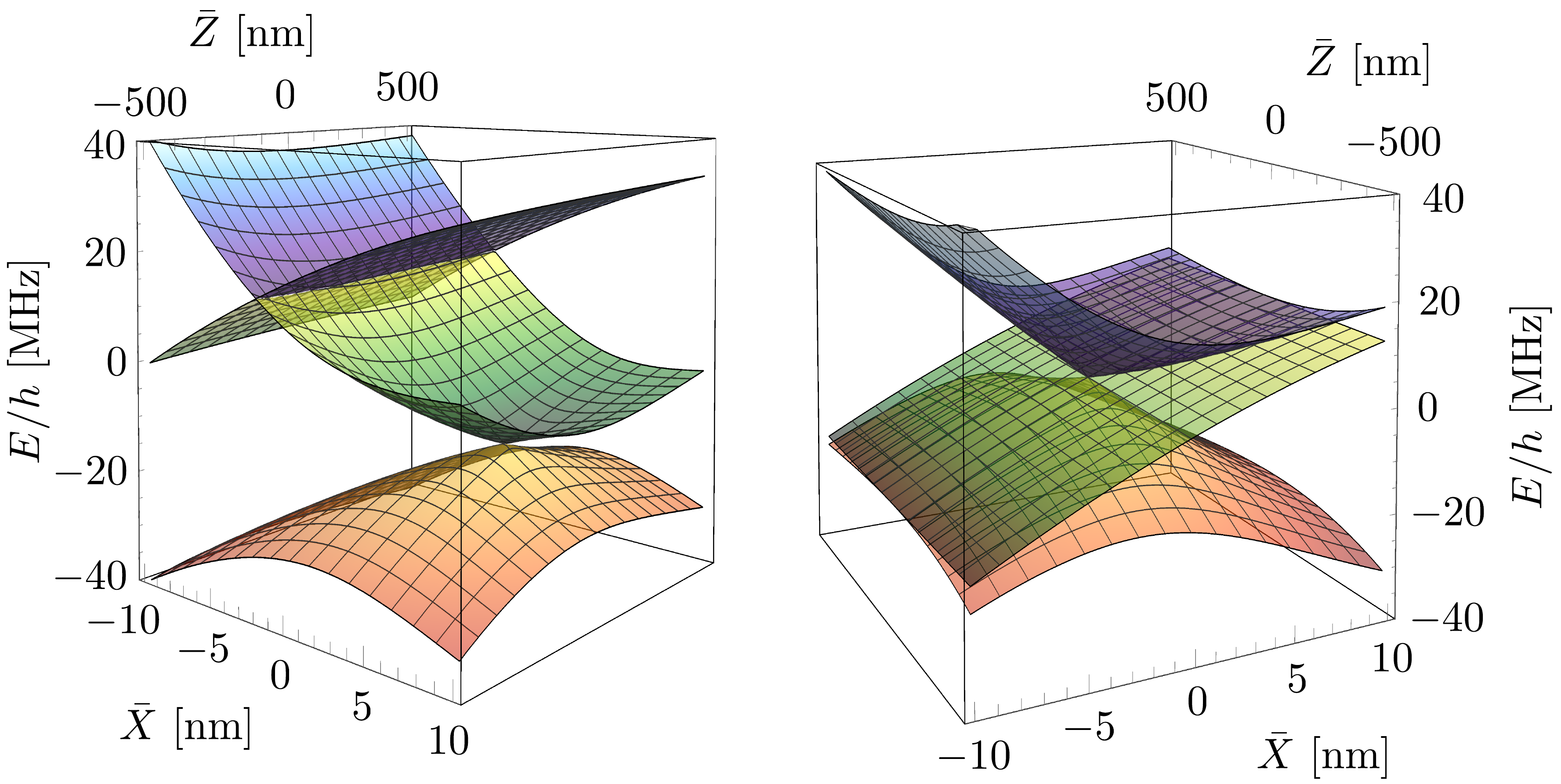}
	\caption{{\bf Multiple-PES CI in a three-ion setup.} Plot of the eigenvalues of $ H_\mathrm{spin}^{1} $ in the $ \bar{X}-\bar{Z} $ plane.
		In (a), the parameters of the MW dressed potential described previously are chosen in such a way that $ V_\mathrm{ex}(R_{12}^0)=V_\mathrm{ex}(R_{23}^0)=0 $. This corresponds to setting $ \Omega_{\mathrm{MW}}=2\pi\times40 $ MHz and $ \Delta_{\mathrm{MW}}=2\pi\times20.9  $ MHz. A CI between the lower (red surface) and the intermediate (green surface) PESs emerges at $ (\bar{X},\bar{Z})\approx(5,0) \ \mathrm{nm}$ while a seam of intersections between the intermediate and the upper (blue surface) PESs can be seen along the $ \bar{Z} $ coordinate for $ \bar{X}\approx 3 \ \mathrm{nm} $.
		In (b), we show the eigenvalues of $ H_\mathrm{spin}^1 $ in the presence of a modified MW dressed potential with vanishing next-to-nearest neighbor exchange interactions. In this case, the CI is shifted to $ (\bar{X},\bar{Z})\approx(0,0) \ \mathrm{nm}$, where all the three PESs are degenerate. Moreover, the intermediate PES forms a seam of intersections with the other two PESs along the $ \bar{Z}=0 $ axis. In both panels, unmentioned parameters are as in Fig.~(2) of the main text. \label{SM:fig:3ionsCI}}
\end{figure*}

The eigenvalues of $ H_\mathrm{spin}^1 $ in the $ \bar{X}-\bar{Z} $ plane are shown in Fig.~\ref{SM:fig:3ionsCI}(a). A CI between the lowest and the intermediate PESs is well visible at $ (\bar{X}, \bar{Z})\approx(5,0) \ \mathrm{nm} $, while the intermediate and upper PESs form a seam of intersections for $ \bar{X}<0 $. In order to simulate the quantum dynamics of the three-ion system in this complex energy landscape one would have to include all the three PESs, which would require a huge amount of computational resources. Indeed, the typical extension and length-scale of the nuclear wavefunction and motion, respectively, are of the order of few nanometers. They are therefore compatible with the distance between the various crossings shown in Fig.~\ref{SM:fig:3ionsCI}(a). On the other hand, a setup of three trapped Rydberg ions provides a state-of-the-art quantum simulator in which the dynamics around CIs involving multiple PESs can be investigated experimentally. 

Furthermore, in analogy with the two-ion setup, the position of the CI shown in Fig.~\ref{SM:fig:3ionsCI}(a) can be controlled via the MW dressed exchange interaction potential. As an example, by tailoring the bichromatic MW field described above in this Supplementary Material, one can modify the strength of the exchange interactions between next-to-nearest neighbors (NNN). In the limit of vanishing NNN interactions, i.e. setting $ V_\mathrm{ex}(R_{13})=0 $, one obtains the PESs shown in Fig.~\ref{SM:fig:3ionsCI}(b). In this case, the CI occurs at $ (\bar{X}, \bar{Z})=\bm{0} $. Here, the three PESs are degenerate and, hence, equally involved in the CI-induced dynamics. Also, notice that the intermediate PES form a seam of intersection along the $ \bar{Z}=0 $ axis with the upper (lower) PES for $ \bar{X}>0 $ ($ \bar{X}<0 $).

\end{document}